\documentclass[11pt]{amsart}
\usepackage{geometry}                
\usepackage{graphicx}
\usepackage{amssymb}
\usepackage{epstopdf}
\usepackage{lipsum}
\usepackage{scrextend}
\title{Conventional Quantum Theory Does Not Support A Coherent Relational Account}
                                     
\begin{document}
\maketitle

\centerline{Ruth E. Kastner} \smallskip
\centerline{University of Maryland, College Park}\smallskip
\centerline{September 23, 2024}\smallskip

ABSTRACT.   Quantum theory in its conventional formulation is notoriously subject to various measurement-related paradoxes, as exemplified by the ``Schr{\"o}dinger's Cat'' and ``Wigner's Friend'' thought experiments. It has been shown, for example by Frauchiger and Renner, that nested measurements such as those occurring in the Wigner's Friend experiment can lead to inconsistencies concerning the putative outcomes of measurements. Such inconsistencies are commonly presumed to remain private and incommensurable, but this is not the case. A counterexample, in which the inconsistencies can be revealed among the observers, is reviewed. The implications for a recent attempt to shield Relational Quantum Mechanics from such inconsistencies are considered, and it is concluded that the attempt is not successful. Further implications for the state of the debate concerning the viability of quantum theory in its various formulations are discussed.

 \medskip

1. Introduction and Background\medskip

	This paper reviews the challenge to conventional quantum theory from inconsistencies arising from its prediction of macroscopic superpositions such as Schr{\"o}dinger's Cat and Wigner's Friend. I focus on an interpretational approach first proposed by Rovelli (1996) called ``Relational Quantum Mechanics''(henceforth RQM). RQM is revisited with modifications in Adlam and Rovelli 2022. The latter paper is the subject of the present critique.
	
	 First, let us clarify what is meant by ``conventional quantum theory'' for purposes of the present work. This characterization is meant to encompass all formulations of quantum theory that provide a quantitative, physical account {\it only} of unitarily evolving processes, i.e., those described by the Schr{\"o}dinger equation. In view of that feature, we can for greater clarity use the term {\it Unitary Quantum Theory} to denote such formulations (which is consistent with terminology in Adlam and Rovelli 2022). This term applies not only to Everettian approaches (which deny non-unitary collapse or reduction), but also to so-called ``textbook quantum theory.'' Although the latter makes use of a projection postulate involving non-unitary collapse, that is traditionally invoked as a purely calculational step for empirical adequacy and utility. The postulate provides no quantitative physical account for a ``measurement'' instantiating non-unitary evolution. Thus, quantum theory's physical (as opposed to mathematical/formal) content is taken as limited to unitary evolution in both Everettian and textbook formulations. Under this broad class of formulations, one can be agnostic about whether the projection postulate should be part of quantum theory, but one assumes that the theory ``really'' has only unitary evolution in a physically quantifiable sense. And that is the prevailing convention.
	 	 
	 The projection postulate was termed {\it Process 1} by Von Neumann (1932), and consists of two distinct mathematical stages. The first is the transition of a pure quantum state, represented by a Hilbert space vector, to a proper mixed state, represented by a density operator on Hilbert space. The second stage is the reduction or collapse to a pure state eigenvector of the measured observable. However, as noted above, in textbook quantum theory, Process 1 has no quantitative physical account that specifies how such a transition, lying outside the  Schr{\"o}dinger evolution, occurs. (The unitary Schr{\"o}dinger evolution was called {\it Process 2} by Von Neumann.)
	 
	  The textbook theory leads to the so-called ``Shifty Split'' in which Schr{\"o}dinger unitary evolution is applied to some subset of all systems in the universe\textemdash the system of interest\textemdash while everything else is taken as outside the domain of applicability of quantum theory (showing again that the theory is assumed to be unitary-only in actual physical content). This places the ``observer'' outside the theory, yet the theory itself cannot specify what constitutes an ``observer'' as a physical system.  Thus the split becomes an {\it ad hoc} device that is applied without a well-defined physical justification. This situation is an aspect of the measurement problem, i.e., the fact that the theory cannot quantitatively account for the occurrence of outcomes of measurements describable by eigenstates of the measured observable, since the latter are in general not arrived at through unitary evolution.\footnote{Everettian approaches often claim to resolve the measurement problem, but they depend strongly on decoherence arguments which ultimately involve taking improper mixtures as providing justification for outcomes, an arguably illicit step. This approach has been critiqued in Kastner (2014), which raises a circularity objection, and Kastner (2022), which raises an inconsistency objection.}
 	   
	 The measurement problem of Unitary Quantum Theory is notoriously exemplified by such paradoxical thought experiments as Schr{\"o}dinger's Cat and Wigner's Friend, which involve ``absurd'' macroscopic superpositions (the characterization in Frauchiger and Renner 2018, and clearly Schr{\"o}dinger's intent). Since the theory (even if utilizing a ``projection postulate'') denies any real physical non-unitarity, it must treat observed determinate outcomes as brute empirical facts with no physical explanation within the theory. More strongly, observed outcomes fail to be predicted by the theory, as the macroscopic superposition paradoxes illustrate. This point is granted by Adlam and Rovelli 2022, who say (in the context of a Wigner's Friend experiment): \smallskip
	
	 {\small``Unitary quantum mechanics does not provide any mechanism for a single measurement outcome to be selected and actualised for Alice in the first place, so it certainly cannot tell us anything about the relationship between her outcome and Bob’s outcome - this is a question which lies entirely outside the unitary part of the theory...'' (Adlam and Rovelli 2022)} \smallskip
	
	The locution ``unitary part of the theory'' might seem to suggest that there is another ``part of the theory,'' but, as discussed above, that merely leaves room for an {\it ad hoc} application of a  collapse postulate, or other means of stipulating an outcome, which is subject to the Shifty Split between the system to which quantum theory is taken as applying and some putative ``outside observer.'' This situation is what gives rise to the Wigner's Friend and related inconsistencies such as that of the Frauchiger-Renner Paradox (Frauchiger and Renner, 2018). Of course, there are many attempts to deal with the measurement problem and resulting inconsistencies arising from Unitary Quantum Theory. Adlam and Rovelli 2022 is one of those attempts; I argue in what follows that it fails.

	In contrast to Unitary Quantum Theory, formulations involving objective reduction or collapse (henceforth OR) are not subject to the measurement problem and thus do not lead to macroscopic superpositions such as that of Bob and his lab in the Wigner's Friend paradox.\footnote{Some examples are the gravitational collapse model of Penrose 2000, the Ghirardi-Rimini-Weber (1986) model, and the Transactional Interpretation (Cramer 1986, most recent relativistic version in Kastner 2022). The latter is distinct from the two former models in that it does not modify the Schr{\"o}dinger equation but uses a different theory of fields that naturally leads to a non-unitary dynamics.} This is the case regardless of one's opinion of the theoretical or practical adequacy of such theories, which is a separate issue. Interestingly, one of the authors of Adlam and Rovelli 2022 (EA) does acknowledge that a particular OR formulation solves the measurement problem in a footnote to a paper entitled ``Do We Have {\it Any} Viable Solution to the Measurement Problem?", where she says: 
	
	{\small ``...the transactional interpretation...is a solution to the measurement problem which has the property of being retrocausal.''} (Adlam 2023).\footnote{Adlam's references are out of date, however; in addition to Cramer (1986), a more recent, fully relativistic account is Kastner (2022); see also Kastner (2018) in regard to TI's solution to the measurement problem.} 
	
	(This comment was made as an aside in the context of properties a particular formulation or interpretation might have. Oddly, however, although acknowledged by the author as a solution to the measurement problem in the quoted footnote, the transactional interpretation is not mentioned anywhere in the body of the paper, despite its title\textemdash even with the emphasis on ``Any'').
	
	Thus, at the risk of repetition (for some emphasis seems necessary), OR formulations eliminate the measurement problem so that such paradoxes as Schr{\"o}dinger's Cat and Wigner's Friend do not arise in the first place. It is important to keep this point\textemdash that the notorious measurement-related paradoxes do not arise under OR formulations\textemdash firmly in mind, since it appears to have been somehow forgotten in Adlam and Rovelli 2022. To elaborate a bit, although it should be obvious: OR formulations involve genuine physical non-unitarity in some form, such that the measurement process is defined at the appropriate micro-level and an interaction constituting a `measurement' according to that theory results in an objectively existing outcome such that the measured system is accurately described by the corresponding eigenstate. This means that an OR formulation (such as the transactional interpretation, TI and its relativistic version RTI; cf. Kastner 2022) would dictate that Schr{\"o}dinger's Cat is not in a macroscopic superposition, since there definitely is or is not an atomic decay in the box at some particular time (cf. Kastner and Cramer 2018). It would similarly dictate that Bob and his lab are not in a macroscopic superposition under the envisioned circumstances of Bob's measurement. Thus Wigner, under an OR theory, will accordingly predict that his friend Bob will actually get an outcome corresponding to an eigenstate for his measured system, and Wigner will assign the appropriate proper mixed state to Bob's system if Wigner does not currently know that outcome. Wigner will {\it not} describe Bob and his lab by a macroscopic superposition, for the OR theory would not predict that (as discussed above). Only Unitary Quantum Theory predicts that. As noted above, whether or not one happens to like or accept or believe a particular OR theory to be fully adequate is irrelevant here: such a theory specifically says that measurements involve collapse and predicts that superpositions therefore do not persist beyond that point. And indeed Adlam acknowledges in (2023) that one such OR theory  (the Transactional Interpretation)``solves the measurement problem,'' as noted above. So one cannot purport to be applying an OR theory while assuming the existence of macroscopic superpositions precluded by OR theories. Yet, as we will see in what follows, Adlam and Rovelli 2022 do just that. \vspace{3mm}

2. Critique \vspace{3mm}

The above quote from Adlam and Rovelli 2022, namely:
	  {\small ``Unitary quantum mechanics does not provide any mechanism for a single measurement outcome to be selected and actualised for Alice in the first place....''} illustrates the conventional assumption of Unitary Quantum Theory that the actual physical content of quantum theory must be restricted to unitary evolution, and therefore that measurement outcomes are not accounted for by the theory but are brute matters of fact for which some `interpretational' account is needed. However, before one even gets to issues of consistency or coherence, it is worth noting that RQM simply helps itself to the observed fact, admitted by the authors as not at all accounted for by Unitary Quantum Theory, that measurements have outcomes\textemdash even if only `relative' to some observer or another. This is not a matter of the theory being `interpreted' but rather an {\it ad hoc} patching of its key, acknowledged lacuna, since the presumed unitary-only physical evolution of the conventional theory itself {\it licenses no outcomes, `relative' or otherwise}. In this sense, RQM simply replaces the usual {\it ad hoc} projection postulate with a slightly different {\it ad hoc} quasi-projection postulate: namely, ``outcomes occur relative to observers.'' It is important to recognize that the latter is not a physical postulate that serves to {\it explain} empirical data, but instead simply appropriates by fiat the very empirical data that its underlying theory admittedly fails to explain. RQM thus postulates the explanandum, and that makes RQM vacuous as an explanans.
	  	
	In contrast,  OR theories, as noted above, provide for the actualization of an outcome under specific physical circumstances involving some form of non-unitarity, and are not observer-dependent. Thus, there is a well-defined point at which collapse/reduction occurs, yielding an outcome as an invariant physical event, and any observer describes the situation accordingly. This precludes  the Wigner's Friend and related ``nested observer'' scenarios. However, Unitary Quantum Theory remains subject to such scenarios, which raise the possibility, made explicit in the Frauchiger-Renner paradox, that the `inner observer' could be found to make a `wrong' prediction. The latter circumstance refutes the central postulate of RQM, stated in Laudisa and Rovelli (2021) as follows: \medskip
	
	{\small ``The physical assumption at the basis of RQM is the following postulate:  The probability distribution for (future) values of variables relative to $S$ depend on (past) values of variables relative to $S$ but not on (past) values of variables relative to another system $S^\prime$.''}\footnote{Laudisa and Rovelli (2021). It should be noted, however, that this is not a ``physical assumption.'' It is a calculational, instrumentalist postulate that says nothing about any underlying physics. In contrast, Boltzmann's assumption that the large-scale laws of thermodynamics were based on the microscopic behaviors of atoms was a physical postulate.}  \medskip
		
	The form of RQM reviewed and slightly revised in Adlam and Rovelli 2022\textemdash call it RQM22\textemdash thus abandons the central postulate of RQM (although the present author has been unable to find any admission on the part of RQM advocates of the fact that its core assumption is violated by the Unitary Quantum Theory that they purport to be interpreting). The original theoretically contradicted postulate is now replaced by the following five postulates: \medskip
	
{\small \noindent 1. Relative facts: Events, or facts, can happen relative to any physical system. No hidden variables: unitary quantum mechanics is complete \newline
2. Relations are intrinsic: the relation between any two systems A and B is independent of anything that happens outside these systems’ perspectives \newline
3. Relativity of comparisons: it is meaningless to compare the accounts relative to any two systems except by invoking a third system relative to which the comparison is made. \newline
4. Measurement: an interaction between two systems results in a correlation within the interactions between these two systems and a third one; i.e. with respect to a third system W, the interaction between the two systems S and F is described by a unitary evolution that potentially entangles the quantum states of S and F. \newline
5. Internally consistent descriptions: In a scenario where F measures S, and W also measures S in the same basis, and W then interacts with F to ‘check the reading’ of a pointer variable (i.e. by measuring F in the appropriate ‘pointer basis’), the two values found are in agreement. 
(quoted from Adlam and Rovelli 2022).} \medskip

These postulates introduce undefined terms such as the `fact' and `event' of Postulate 1. As mentioned above, if these are supposed to denote measurement outcomes, the authors help themselves to the occurrence of outcomes `for observers' despite the fact, acknowledged by them, that {\small  ``Unitary quantum mechanics does not provide any mechanism for a single measurement outcome to be selected and actualised for Alice in the first place...''.}\footnote{The fact that RQM helps itself to the notion of ``measurement'' and ``measurement outcomes'' not licensed by Unitary Quantum Theory is demonstrated in Rovelli (1998), which dodges the issue entirely by saying ``For some reason, at some point we have to (or we can) replace the pure state $\Psi(T )$ with a mixed state.'' It is then simply stipulated without justification that the system acquires some eigenvalue of the observable in question.}  While the authors go on to talk about ``flashes'' as a matter of purported ontology, under Unitary Quantum Theory these {\it ad hoc} entities can never constitute the ``observer-independent events'' offered, as we will see in what follows. In addition, Postulate 3's purpose seems unclear. If the intent is to rule out an observer-independent account, one might wonder why that would be necessary in view of the authors' claim that their interpretation yields observer-independent facts.\footnote {Deeming certain questions ``meaningless'' can serve to evade legitimate questions; in particular, comparison of accounts would seem to be meaningful if one is purporting to provide a physical and even `observer-independent' ontology of the world, which Adlam and Rovelli are doing.}

As an aside, the present author certainly agrees with the authors that one should aim for a fully relativistic interpretation, one which takes into account that spacetime could be emergent from the quantum level and thus is not assumed as a fixed background or container for events, as in their remark: ``there is an alternative view which suggests that in fact, spacetime should be understood to emerge from a background of quantum events.''  Indeed a comprehensive account of just how this could work is provided in Schlatter and Kastner (2023). The quote from Born is also apt:  ``One should not transfer the concept of space-time as a four-dimensional continuum from the macroscopic world of common experience to the atomistic world; manifestly the latter requires a different type of manifold.'' However, these considerations do not save RQM, nor its recent reworking in Adlam and Rovelli 2022, from the problems discussed herein.

	The authors review the Wigner's Friend scenario to which Unitary Quantum Theory is subject, in which an `inner observer' (Bob) performs a measurement M, finds an outcome and attributes the relevant eigenstate to his system, while Wigner (the `outer observer') describes Bob with a coherent superposition, so that the two disagree on what quantum theory predicts. They claim that this disagreement cannot be corroborated in any way: {\small``Wigner and Bob simply live within incommensurate realities now, and no attempt to reach across and bridge the gap can possibly succeed.}''. (Adlam and Rovelli 2022)
	
	 The authors then focus on this alleged situation (which is falsified by a counterexample to be discussed in Section 2b) as something to be improved upon and to that end suggest replacing Postulate 4 with Postulate 4.1 involving ``cross-perspective links'': \medskip
	
	 {\small ``Definition 4.1. Cross-perspective links: In a scenario where some observer Alice measures a variable V of a system S, then provided that Alice does not undergo any interactions which destroy the information about V stored in Alice’s physical variables, if Bob subsequently measures the physical variable representing Alice’s information about the variable V, then Bob’s measurement result will match Alice’s measurement result.''} (Adlam and Rovelli 2022) \medskip

	However, 4.1 does not succeed in remedying the essential inconsistency faced by Unitary Quantum Theory, as we will see in what follows. We now elucidate four consequential errors made by the authors:
	
\medskip		
\noindent Error 1: Failure to actually apply an OR theory while claiming to apply it\newline
Error 2: Failure to acknowledge empirically consequential nature of inconsistent predictions\newline
Error 3: Unsupported claim that Wigner's measurement `destroys' Bob's outcome\newline
Error 4: Infinite regress involving `superobservers' nullifies consistency claim\newline

\medskip
	I now discuss each of these errors in turn.\medskip

{\it 2a. Error 1: Failure to apply an OR theory while claiming to apply it}\medskip

The authors review the Wigner's Friend scenario and then comment:\smallskip

{\small ``Now, if we think that measurement M must have a definite outcome, then clearly we must conclude that either Wigner or Bob is wrong. Wigner is wrong according to any interpretation which tells us that people cannot be in superpositions - e.g. gravitational collapse interpretations and spontaneous collapse interpretations - because those interpretations affirm that there will be an actual or effective collapse of the wavefunction at around the time of Bob’s measurement so that Bob will no longer be in a coherent superposition by the time of Wigner’s measurement.''}  \smallskip

	This is a very peculiar statement. The assertion ``Wigner is wrong'' arises only from the authors' refusal to allow Wigner to apply the very OR formulation they claim to be applying. As discussed above, OR instructs Wigner {\it not} to describe Bob by a macroscopic superposition but by a proper mixed state. The authors even admit this: ``those [OR] interpretations affirm that there will be an actual or effective collapse of the wavefunction at around the time of Bob’s measurement so that Bob will no longer be in a coherent superposition by the time of Wigner’s measurement."  So why do they assume that Wigner will describe Bob by a coherent superposition dictated not by OR but by Unitary Quantum Theory? Perhaps more to the point, why do the authors ignore the fact, just stated by them, that the OR theory dictates that ``Bob will no longer be in a coherent superposition by the time of Wigner’s measurement''? 
	
	The fact that OR theories preclude the Wigner's Friend-type paradoxes is acknowledged by Baumann and Brukner (2019, 3), who say:
	
	{\small ``[OR]  theories postulate a modification of standard quantum mechanics, such as spontaneous [7] and gravity-induced [4, 8] collapse models, which become significant at the macroscopic scale. These models could exclude superpositions of observers as required in Wigner’s friend-type experiments and hence the protocol cannot be run.'' }\footnote{Baumann and Brukner go on to say ``In our view this is the most radical position as it would give rise to new physics.'' However, the widespread assumption that OR theories necessarily depart from the predictions of quantum theory is incorrect, since at least one OR-type theory, the Transactional Interpretation, not only does {\it not} depart from the Born Rule but physically derives it. So any ``new physics'' of TI is confined to the very old but overlooked physics of the direct-action theory of fields, which yields an empirically equivalent form of quantum theory that also accounts for ``measurement'' (cf. Kastner 2022, Chapter 5).} \smallskip
	
 Now, if one thought a particular OR formulation failed to accomplish actual reduction at a macroscopic level, one could quibble with that. But that is not what the authors have done. They admit that an OR theory precludes macroscopic superpositions. Yet they unaccountably persist in assuming that the Wigner's Friend scenario still applies, falsely implying that the OR theory does not give a satisfying resolution of a `paradox' to which it is not even subject, while their proposed interpretation supposedly does.  \medskip

{\it 2b. Error 2: Failure to acknowledge empirically consequential nature of  inconsistent predictions}\smallskip

	The inconsistencies that arise in the MP-generated paradoxes involve supposedly observed facts or predictions that are in conflict among different observers. Many discussions of these inconsistencies assume that these putative observed facts cannot be compared among observers of different ``nesting'' levels, and thus constitute ``incommensurate realities." This  assumption exemplified in Adlam and Rovelli 2022 in their comment:
	
{\small ``Wigner and Bob simply live within incommensurate realities now, and no attempt to reach across and bridge the gap can possibly succeed.''}

The authors then attempt to remedy that asserted situation with what they term `cross-perspective links' as proposed in Postulate 4.1. However, this revision does nothing to remedy the fact that the `inner observer' can indeed find that his prediction deviates from that of the `outer observer' (according to Unitary Quantum Theory). This is demonstrated in an extant example discussed in Kastner 2020, 2021, 2022, and Baumann and Brukner 2019, which we now review (closely following the account in Kastner, 2022). The example, a variant of the Wigner's Friend experiment, demonstrates that attributing outcomes to entangled subsystems, even if designated as only ``relative'' to the subsystem or involving only ``subjective collapse,'' (as discussed by Baumann and Wolf, 2018), leads to overt empirical inconsistency as opposed to ``hidden'' and ``incommensurate'' inconsistencies as commonly assumed.\footnote{Deutsch (1985) considers a similar scenario, but offers it as an empirical comparison between the Everettian picture and the Copenhagen interpretation. Here I discuss implications of this type of empirical disagreement that do not yet seem to have been fully appreciated.} 

   First, recall that Unitary Quantum Theory has no well-defined criterion for what sort of system counts as a ``measurement device''  to which  an outcome could be attributed, nor for what sort of system counts as an ``observer'' who could be taken as acquiring knowledge of an outcome. This leads to the common treatment, as for example in Proietti {\it et al} (2019), in which a microscopic quantum system such as an atom, molecule, or photon is taken as an ``observer'' whose correlation with another degree of freedom can (when desired) constitute a ``measurement'' yielding an outcome. With that background, consider a Wigner's Friend experiment in which the two parties, labeled $W$ and $F$, are instantiated by degrees of freedom subject to 2D state spaces, such as spin-$\frac{1}{2}$ atoms. As usual, $F$ is taken as an ``inner observer'' and $W$ as an ``outer observer.'' $F$ comprises 2 degrees of freedom: $B$, acting as a pointer/memory, and $C$, for communication.  $C$'s possible states are `ground' and `excited,' where the latter would be triggered by a photon signal from the outer observer $W$.
   
       Let us suppose that $F$ is measuring  another spin-$\frac{1}{2}$ system $A$, prepared in an equal superposition of states `up' and `down' along a direction $Z$, via a Stern-Gerlach device. One can think of $F$ as a micro-detector whose $B$ states become correlated with $A$ via an appropriate interaction Hamiltonian. $C$ remains in its initial unexcited state $|0\rangle$ at this stage. 
       
       Now, based on the conventional approach of attributing an outcome to a system based on the establishment of a suitable interaction Hamiltonian as above, according to $F$ after the ``measurement''  of $A$ by $B$, $A$ the relevant degrees of freedom are either in the state:  
      
      $$ |\Psi\uparrow\rangle_{FA} =   |\uparrow\rangle_A  \odot |\uparrow\rangle_B \odot |0\rangle_C\eqno(1)$$
      or
       $$ |\Psi\downarrow\rangle_{FA} =  |\downarrow\rangle_A  \odot |\downarrow\rangle_B \odot |0\rangle_C\eqno(2)$$
       \smallskip
       
\noindent with equal probability--i.e., they are in a mixed state.  On the other hand, according to $W$,  $A$ and $F$ end up in a pure Bell state with $F$'s communication degree of freedom $C$ along for the ride:

        $$ |\Psi \rangle_{FA} = |\Phi^+ \rangle |0 \rangle_C = \frac{1}{\sqrt 2} \bigl ( |\uparrow\rangle_A  |\uparrow\rangle_B
        + |\downarrow\rangle_A  |\downarrow\rangle_B \bigl ) \odot |0\rangle_C\eqno(3)$$
        
Now, let $W$ subject the $B+A$ system  to a measurement of the Bell observable for which the state $|\Phi^+\rangle$  is an eigenstate, where only the outcome $\Phi^+$ triggers the emission of a photon to $F$ to excite his communication degree of freedom $C$.\footnote{The Bell observable is not uniquely required. Alternatively, one could arrange for $F$'s pointer $B$ to be anti-correlated with $A$'s putative outcomes, and have $W$ conduct a measurement of the total spin of the system. This would reveal the inconsistency since $W$ predicts that the total spin $J$ must always be found to be 1, while according to F, W has a 50\% chance of finding the result $J=0$, for the singlet state.}  According to $W$,  $F$ should receive that photon for every run of the experiment.  However, according to $F$, his probability of receiving the photon is only 1/2  in view of his description of the situation by the mixed state (1,2). This makes the inconsistency manifest.

	We should note that the "basic postulate of RQM" (Rovelli 1986), repeated by Laudisa and Rovelli 2021)  is refuted by this example: 
	
{\small ``The physical assumption at the basis of RQM is the following postulate:  The probability distribution for (future) values of variables relative to $S$ depend on (past) values of variables relative to $S$ but not on (past) values of variables relative to another system $S^\prime$.''}\smallskip

Thus, RQM asserts that $F$ should be able to predict correctly the future values of its own local system(s) by reference only to its local past outcomes, while its presumed unitary entanglement with the result of the world continues. But in the counterexample (presuming that $W$ is correct, as Adlam and Rovelli assert),  $F$ {\it cannot} correctly predict the probability of a future event relative to itself---i.e., the probability of its degree of freedom $C$ becoming excited---based on its own past outcomes. This probability apparently depends instead on another system outside ($W$), contrary to the central postulate of RQM. At this point, RQM is effectively dead. The version offered in Adlam and Rovelli 2022, RQM22, is a mere shadow of its former self. Yet even that shadow fails to survive, as we shall see in what follows.  \medskip

{\it 2c. Unsupported claim about `destroyed information' } \smallskip

The authors first focus on alleged incommensurability of the inner and outer observers with respect to a commuting observable, i.e. (with Alice playing the role of F and Bob playing the role of W). They say:\smallskip
 
{\small ``Now suppose Bob measures S in the same basis as Alice’s measurement, and hence he obtains a measurement outcome MBS which he will interpret as providing information about the result of Alice’s measurement on S. Suppose that Bob also ‘measures’ Alice herself and obtains a measurement outcome MBA for the value of of some pointer variable which is supposed to be a record of her measurement result - for example, he could simply ask her what her measurement result was. So in this scenario we have three different measurement outcomes MA,MBS,MBA all supposedly providing information about the value of the same variable.''} (Adlam and Rovelli 2022)

The authors then lament the alleged fact that these results are incommensurable and go on to replace Postulate 4 with  Postulate 4.1 concerning ``Cross-perspective Links'', which we repeat here for convenience:\smallskip
 
 {\small Definition 4.1. Cross-perspective links: In a scenario where some observer Alice measures a variable V of a system S, then provided that Alice does not undergo any interactions which destroy the information about V stored in Alice’s physical variables, if Bob subsequently measures the physical variable representing Alice’s information about the variable V, then Bob’s measurement result will match Alice’s measurement result.''} (Adlam and Rovelli 2022 \smallskip

  The provision about destroyed information does a lot of `heavy lifting' in the authors'  subsequent attempt to remove the threat to RQM of the Wigner's Friend paradox (as we will discuss further below). However, the system in the counterexample has a degree of freedom C that is not subject to any interaction that would `destroy' its information, and F makes a specific prediction about the probability of a state change in C that disagrees with the probability predicted by W. If one takes the `correct' description as that of F being in a pure state superposition, then F must find out he is wrong, violating the central postulate of RQM as noted above. Does RQM22 do any better?  The authors offer the following:

{\small ``But when postulate four is replaced with cross-perspective links, we have a different story. In this approach, Bob performs his measurement and the system he measures takes on a definite value relative to him, and then perhaps he will be tempted to reason according to the steps set out in ref [42] to conclude that the measurement M will definitely have outcome 0. But relative to Wigner, Bob is still in a superposition state and thus Wigner is able to perform a measurement on Bob in which the value of the variable measured by M takes on the value 1 relative to Wigner. Moreover, although this value is relativized to Wigner, it is also an {\bf observer-independent fact} [emphasis added] in the sense that other observers can find out about it by measuring or indeed just asking Wigner - for example, if Bob gets out of the box and asks Wigner what has transpired, Wigner will tell him that the measurement had outcome 1 and Bob will perceive Wigner saying that the measurement had outcome 1. So Bob is wrong, although since the measurement on Bob will destroy the information stored in Bob’s physical variables about the outcome of his measurement and hence also any inferences he made on the basis of that measurement, Bob will necessarily have no memory of making a wrong prediction, so no direct contradiction will ever arise.''}\smallskip

	However, in the above counterexample, according to Unitary Quantum Theory, the inner observer will indeed find that he is `wrong' just by checking his degree of freedom C. There is no `box' to step out of, nor any mechanism for wiping his memory.  The assumption that the inner observer's memory will be wiped is merely an article of faith that is falsified by the counterexample. RQM22 fares no better than RQM, falling victim to the acknowledged inability of Unitary Quantum Theory to account for determinate outcomes.  \medskip

{\it 2d. Infinite regress involving `superobservers' nullifies consistency claim} \medskip

As noted above, the authors assert that Wigner's outcome is an `observer-independent fact' (boldface sentence). Yet Unitary Quantum Theory fails to license this assumption, since one could always have another `outside observer' Wigner2 who (according to Unitary Quantum Theory) must describe Wigner as being in a coherent superposition after his measurement\textemdash since according to Unitary Quantum Theory, no observer is ever correct in thinking that any collapse to an eigenvalue actually occurred. This means that Wigner's outcome cannot be an `observer independent fact' (at least as a piece of information supporting a correct prediction) as asserted in the quoted excerpt. This point reveals the fallacy in the following argumentation by the authors:	\smallskip

{\small ``Note also that Bob’s erroneous prediction arises only because he assumes that there is some kind of collapse when he performs his measurement. If instead he uses the version of RQM that we have suggested in this article, then he will correctly predict that Wigner could still get the result 1, since he will be aware that although his observation has caused him to update his own relative state, he remains in a superposition state relative to Wigner (since that state describes the relational [sic] between Wigner and Bob, not the absolute state of Bob). So in fact no contradictory predictions will arise as long as Bob and Wigner are consistent about employing the same interpretation of quantum mechanics.'' } (Adlam and Rovelli 2022) \smallskip

The infinite regress  becomes obvious here if we simply replace Bob by Wigner and Wigner by Wigner2, who measures some other observable $X$: \smallskip

{\small \bf  ``Note also that Wigner's erroneous prediction arises only because he assumes that there is some kind of collapse when he performs his measurement. If instead he uses the version of RQM that we have suggested in this article, then he will correctly predict that Wigner2 could still get the result $x$, since he will be aware that although his observation has caused him to update his own relative state, he remains in a superposition state relative to Wigner2 (since that state describes the relational [sic] between Wigner2 and Wigner, not the absolute state of Wigner). So in fact no contradictory predictions will arise as long as Wigner and Wigner2 are consistent about employing the same interpretation of quantum mechanics."} \smallskip 

In other words, the ostensible remedy for the inconsistencies, according to RQM, is to assume that every observer's prediction, based on his putative measurement outcome and assignment of the corresponding eigenstate, is potentially erroneous based on the possible existence of an `outer observer.' Along with this goes the untenable assumption that the inner observer will have no memory of being wrong, which is falsified by the counterexample discussed above. Thus, RQM prescribes that all experimenters are wrong when they attribute outcome eigenstates to their systems, since some outside observer would deny that attribution. But there could never be any correct `outside observer' due to the infinite regress of `outside observers'. And indeed there can be no `observer independent facts' if all observers are mistaken whenever they think that they got an outcome corresponding to an eigenstate of the measured observable.  RQM thus fails to support any observer-independent facts, contrary to the authors' claim, i.e.: 

\smallskip {\small ``...we now regard the pointlike quantum events or `flashes’ as absolute, observer-independent facts about reality, rather than relativizing them to an observer."}  (Adlam and Rovelli 2022) \smallskip

 This simply will not do, since there is no physics given to support such a `quantum event'. If Bob makes a measurement and is taken as finding an outcome, the authors seem to want to take that as a `quantum event' corresponding to a `fact'. But under Unitary Quantum Theory (and RQM and RQM22), that `fact' must be wrong, and so is Wigner's, ad infinitum. In what coherent sense could myriads of wrong impressions possibly count as `observer-independent facts about reality'? 
 
 A further note is in order regarding the approach advocated in Baumann and Brukner (2019) (henceforth BB2019), which seeks to tame the Wigner's Friend-type inconsistencies by recourse to an instrumentalist treatment. They consider essentially the same counterexample discussed in Section 2b, in which an `inner observer'  could (according to Unitary Quantum Theory) find out they were wrong about their outcome-based prediction. BB2019 proceed under the conventional assumption that Unitary Quantum Theory is the correct theory and that the resulting Wigner's Friend-type scenarios are arguments against the existence of `observer-independent facts', as in their remark:
 
 ``In Ref. [2, 3] a combination of two spatially distant Wigner’s-friend setups is used as an argument against the idea of `observer-independent facts' via the violation of a Bell-inequality.''
 
They then offer an instrumentalist approach involving `rational agents' to attempt to accommodate the resulting inconsistencies, in which agents can update their `degree of belief' subject to new information from an `outside observer'. However, BB2019 allow for an `automatic machine' to serve as `Wigner' and to obtain an outcome which it then passes to the Friend (to get around worries about Wigner being dishonest about his outcome). The authors appear not to notice that they have just helped themselves to an observer-independent fact, namely the machine's outcome. In other words, in attributing an outcome to a machine, they tacitly assume that quantum theory does in fact apply to physical systems, not `agents', so that a `fact' can obtain without an agent. Which is it? Are the `facts' corresponding to outcomes observer-dependent or not? The discussion in BB2019 assumes both, despite the fact that they are mutually exclusive.  Moreover, in any case, Unitary Quantum Theory fails to predict any outcomes. If one helps oneself to an outcome despite that, the outcome (or at least any `fact' it might be taken to imply) is subject to being wrong, whether it is attributed to an agent or to a physical system. And the latter has no `degree of belief' to update. The theory\textemdash Unitary Quantum Theory\textemdash that is unaccountably taken as yielding an outcome (or `fact') simply cannot provide a correct one. 

\medskip

3. Conclusion\medskip

	It has been shown that a recent attempt to shield Relational Quantum Mechanics (RQM) from inconsistencies arising from conventional quantum theory's  assumed unitary-only character, and its attendant measurement problem, fails. The attempt founders on an explicit counterexample to the usual claim that discrepancies in predictions of `inner' and `outer' observers in Wigner's Friend-type scenarios remain `hidden' or incommensurable, as well as other errors: (1) failure to actually apply an objective reduction model when the authors claim to be applying it, (2) an unsupportable claim that the outer observer's measurement will `destroy' information available to the inner observer (refuted by the discussed counterexample), and (3) an overlooked infinite regress in which no observer can ever be considered correct because he or she might be an `inner observer'; there is no natural or consistent end-point for the designation of `outer observer.' This situation also negates the authors' claim that their account provides for `observer-independent facts.'  
	
	The above points should also be kept in mind when evaluating the preferred options of Baumann and Brukner (2019), who treat quantum theory in an instrumentalist  fashion, as a tool for making predictions. They refer to the inner observer as being subject to `updating degrees of belief', but as noted above, under that approach there is still in principle an infinite regress of observers whose `observed outcomes' are neither predicted by Unitary Quantum Theory nor can be regarded as facts about reality. Moreover, the latter treatment implicitly smuggles in the observer-independent occurrence of outcomes in assuming that an `automatic machine' could play the part of the outer observer. Yet in that case, the infinite regress still applies, and the `automatic machine' could somehow be `mistaken' about its outcome. Thus, once again, Unitary Quantum Theory is faced with a fatal inconsistency among outcomes that no resort to `updating' can cure, since a machine is not subject to `degrees of belief'. 
	
	In contrast to this unsatisfactory situation facing conventional unitary quantum theory and its putative interpretation RQM, it has been recalled that theories that incorporate objective reduction are not subject to the measurement problem (as acknowledged by one of the authors in a separate publication, Adlam 2023) and thus are not subject to its attendant paradoxes that lead to inconsistencies subject to remedy only by resort to observer-dependent instrumentalism and {\it ad hoc} postulations of outcomes not predicted by the theory itself, and in which all such postulated outcomes are subject to being incorrect.	
\medskip	
	
\noindent Acknowledgments. I am grateful to two anonymous referees for valuable suggestions for improvement of the presentation.

\newpage
 
 References\medskip
 
Adlam, E. (2023). ``Do We Have {\it Any} Viable Solution to the Measurement Problem?'' preprint, https://arxiv.org/pdf/2301.06192.
 
Adlam, E. and Rovelli, C. (2022). ``Information is Physical: Cross-Perspective Links in Relational Quantum Mechanics,'' {\it Philosophy of Physics 1}(1), 4.  \newline Preprint, https://arxiv.org/abs/2203.13342.
 
Baumann, V. and Brukner, C. (2019). ``Wigner's friend as a rational agent,'' preprint, https://arxiv.org/pdf/1901.11274.pdf

Baumann, V. and Wolf, S. (2018). ``On Formalisms and Interpretations,'' {\it Quantum 2}, 99.

Cramer, J.G. (1986). ``The Transactional Interpretation of Quantum Mechanics,''  {\it Rev. Mod. Phys. 58}, 647.
 
Frauchiger, D. and Renner, R. (2018). ``Quantum theory cannot consistently describe the use of itself,'' {\it Nature Communications 9}, Article number: 3711.

Ghirardi, G.C., Rimini, A., and Weber, T. (1986). {\it Phys. Rev. D 34}, 470.
 
Kastner, R. and Cramer, J. (2018). Quantifying Absorption in the Transactional Interpretation. 

Kastner, R. E. (2014). `` `Einselection' of Pointer Observables: The New H-Theorem?'' {\it Stud. Hist. Philos. Mod. Phys. 48}, 56-8.
 
Kastner, R. E.  (2018). ``On the Status of the Measurement Problem: Recalling the Relativistic Transactional Interpretation,'' {\it Int'l Jour. Quan. Foundations 4}, 1:128-141.

Kastner, R. E. (2020). ``Unitary-Only Quantum Theory Cannot Consistently Describe the Use of Itself: On the Frauchiger-Renner Paradox,'' {\it Foundations of Physics 50}, 441-456; https://arxiv.org/abs/2002.01456. 

Kastner, R. E. (2021). ``Unitary Interactions Do Not Yield Outcomes: Attempting to Model `Wigner's Friend','' {\it Found Phys 51}, 89 (2021). doi:10.1007.s10701-021-00492-3; ; https://arxiv.org/abs/2105.01773. 

Kastner, R. E. (2022). {\it The Transactional Interpretation of Quantum Mechanics: A Relativistic Treatment.} Cambridge: Cambridge University Press.

Kastner, R. E. (2023). ``Quantum Theory Needs (and Probably Has) Real Reduction,  Invited contribution to a forthcoming volume in honor of Roger Penrose.   https://arxiv.org/abs/2304.10649.

Laudisa, Federico and Carlo Rovelli (2021). "Relational Quantum Mechanics", The Stanford Encyclopedia of Philosophy (Spring 2021 Edition), Edward N. Zalta (ed.), URL = <https://plato.stanford.edu/archives/spr2021/entries/qm-relational/>. 

Penrose, Roger (2000). ``Wavefunction collapse as a real gravitational effect.'' {\it Mathematical Physics 2000}, 
266–282. World Scientific. doi: 10.1142/9781848160224 0013.
 
Proietti, M. et al (2019). ``Experimental test of local observer independence,'' {\it Science Advances}, 
Vol. 5, no. 9. DOI: 10.1126/sciadv.aaw9832 

Rovelli, Carlo (1996). ``Relational Quantum Mechanics,'' International Journal of Theoretical Physics, 35(8): 1637?1678. doi:10.1007/BF02302261.

Rovelli, Carlo (1998). ‘Incerto tempore, incertisque loci’: Can we compute the exact time  at which a quantum measurement happens? Found. Phys., 28:1031–1043.

Schlatter, A.  and R E Kastner (2023). ``Gravity from Transactions: Fulfilling the Entropic Gravity Program,'' {\it J. Phys. Commun.} 7 065009.
 DOI 10.1088/2399-6528/acd6d7
 
 Von Neumann, John. (1932). {\it  Mathematical Foundations of Quantum Mechanics}. Berlin: Springer. (In German; English edition 1955, Princeton University Press.)

\end{document}